\documentstyle[aps,prl,twocolumn]{revtex}

\begin{document}

\input epsf.sty
\twocolumn[\hsize\textwidth\columnwidth\hsize\csname %
@twocolumnfalse\endcsname

\widetext
\draft

\title
{
Competition between charge/spin-density-wave orders and superconductivity in La$_{1.875}$Ba$_{0.125-x}$Sr$_{x}$CuO$_{4}$
}

\author
{
M. Fujita$^{1}$, H. Goka$^{1}$, K. Yamada$^{1}$, and M. Matsuda$^{2}$
}

\address
{
$^{1}$Institute for Chemical Research, Kyoto University, Gokasho, 
Uji 610-0011, Japan
}

\address
{
$^{2}$Advanced Science Research Center, Japan Atomic Energy Research Institute, Tokai, Ibaraki 319-1195, Japan
}

\date{\today}

\maketitle




\begin{abstract}

{
We have performed a series of elastic neutron scattering measurements on 1/8-hole doped La$_{1.875}$Ba$_{0.125-x}$Sr$_{x}$CuO$_{4}$ single crystals with {\it x} = 0.05, 0.06, 0.075 and 0.085. 
Both charge-density-wave (CDW) and spin-density-wave (SDW) orders are found to develop simultaneously below the structural transition temperature between the low-temperature orthorhombic (LTO) and low-temperature tetragonal (LTT) or low-temperature less-orthorhombic (LTLO) phases. 
In the ground state the CDW order is observed only in the LTT/LTLO phase and drastically degrades towards the LTO boundary. 
The {\it x}-dependence of {\it T}$_{c}$ strongly suggests a direct effect of the CDW order on the suppression of superconductivity. 
Results are discussed in comparison with those from the La$_{1.6-x}$Nd$_{0.4}$Sr$_{x}$CuO$_{4}$ system within the framework of the stripe model. 
}
\end{abstract}


\pacs{PACS numbers: 74.72.Dn, 71.45.Lr, 75.30.Fv, 74.25.Dw} 

\phantom{.}
]
\narrowtext


The interplay between magnetism and superconductivity is a central issue in high-{\it T}$_{c}$ superconductivity ~\cite{Kastner_98}. 
Neutron scattering measurements on the superconducting La$_{2-x}$Sr$_{x}$CuO$_{4}$ (LSCO) system have shown dynamical incommensurate (IC) magnetic correlation \cite{Yoshizawa_88,Birgeneau_88,Cheong91}. 
The linear doping dependence of incommensurability with the superconducting transition temperature ({\it T}$_{c}$) in the under-doped region suggests a relationship between magnetic correlation and superconductivity ~\cite{Yamada_98}. 

On the other hand, it is well known that superconductivity in the La$_{2-x}$M$_{x}$CuO$_{4}$ (M=Ba, Sr) system is anomalously suppressed at the specific hole concentration of {\it x} $\sim $1/8. 
For the La$_{2-x}$Ba$_{x}$CuO$_{4}$ (LBCO) system, this 1/8-anomaly is accompanied by an occurrence of the low-temperature tetragonal (LTT) phase with {\it P}4$_{2}$/{\it ncm} symmetry \cite{Moodenbaugh88,Kumagai88,Axe89}. 
Superconductivity in the low-temperature orthorhombic (LTO) phase of LSCO with {\it B}{\it mab} symmetry is weakly suppressed in comparison with the LBCO system \cite{Takagi89,Kumagai94}. 
Maeno {\it et al}. initiated systematic studies on La$_{1.875}$Ba$_{0.125-x}$Sr$_{x}$CuO$_{4}$ (LBSCO) using polycrystalline samples to clarify the relation between {\it T}$_{c}$ and crystal structure\cite{Maeno91}.
We comprehensively studied the same system using single crystals and revealed a drastic change in {\it x}-dependence of {\it T}$_{c}$ \cite{Fujita_01_2,Goka_unpublish}. 
In the present study we demonstrate a close relation between {\it T}$_{c}$ and crystal structure as shown in Fig.4.

Recently, the IC spin-density-wave (SDW) and the charge-density-wave (CDW) orders were discovered in La$_{1.6-x}$Nd$_{0.4}$Sr$_{x}$CuO$_{4}$ (LNSCO) with {\it x} $\sim$ 1/8. 
These provide important clues for understanding the mechanism of the 1/8-anomaly based on the stripe model \cite{Tranquada95,Tranquada97}. 
In the framework of the stripe model it is recognized that the orders are a manifestation of dynamical spin/charge correlation and that stabilization of the orders competitively induces instability of superconductivity \cite{Baskaran}. 
In the LTT phase stripe-shaped orders parallel or perpendicular to Cu-O bonding are favorably stabilized by the corrugated pattern of the in-plane lattice potential. 
Hence, {\it T}$_{c}$ is suppressed to a greater extent in the LTT phase than in the LTO phase. 
Whether visible signs, such as the anomalous suppression of {\it T}$_{c}$, appear depends on the stability of CDW/SDW orders. 
Therefore investigation of the relationships between the crystal structure, CDW/SDW orders, and superconductivity provides the insight necessary to clarify the mechanism of {\it T}$_{c}$-suppression.

In order to determine the nature of the aforementioned relationships, we performed systematic elastic neutron scattering measurements on a series of LBSCO single crystals. 
As an experimental advantage in this system variation of the Ba/Sr ratio can modify the crystal structure from the LTT to the LTO phase while maintaining a constant carrier concentration ~\cite{Maeno91}. 
Therefore, LBSCO is a favorable compound for investigating the physical properties of a 1/8-doped system in different crystal structures. 
Furthermore, LBSCO system is free from large rare-earth magnetic moments such as Nd spins in LNSCO. 
Our measurements yield two important results: (i) a direct relation between CDW order and suppression of superconductivity, and (ii) the coincident appearance of CDW and SDW orders below the structural transition temperature. 

Single crystals with {\it x} = 0.05, 0.06, 0.075 and 0.085 were grown through a standard traveling-solvent floating-zone method and annealed under oxygen gas flow to minimize oxygen deficiencies. 
Neutron scattering measurements were carried out on the triple-axis spectrometers TOPAN and TAS-1 installed in the JRR-3M reactor at the Japan Atomic Energy Research Institute (JAERI) in Tokai. 
We selected incident neutron energies of {\it E$_i$} = 14.7 meV for TOPAN and 13.7 meV for TAS-1 using the (0 0 2) reflection of a pyrolytic graphite monochromator. 
\vspace{-5mm}
\linebreak
\twocolumn[
\begin{figure}
\vspace{-5mm}
\centerline{\epsfxsize=5.0in\epsfbox{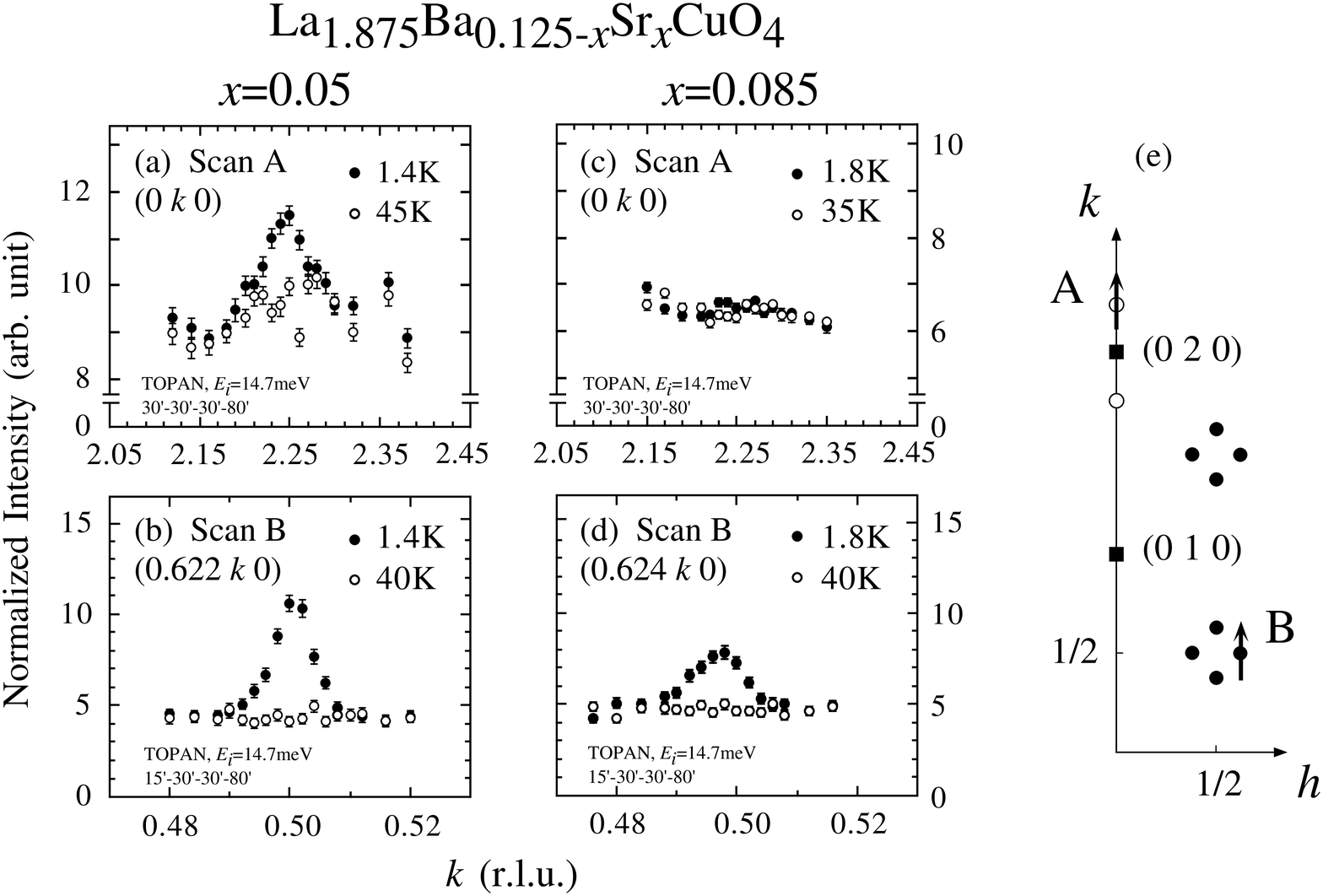}}
\vspace{0.5cm}
\hspace{8mm}
\begin {minipage} {16cm}
\caption
{
IC peaks from CDW and SDW orders in La$_{1.875}$Ba$_{0.125-x}$Sr$_{x}$CuO$_{4}$ with {\it x} = 0.05 ((a) and (b)) and 0.085 ((c) and (d)) measured below {\it T}$_{d2}$ (closed circles) and above {\it T}$_{d2}$ (open circles). Vertical scales are normalized through the sample volumes and counting time. (e) Scan geometry in the ({\it h} {\it k} 0) tetragonal plane. Closed squares show nuclear Bragg peaks; open and closed circles denote nuclear and magnetic IC superlattice peaks, respectively. 
}
\end {minipage}
\end{figure}]
\vspace{0in}
\noindent
Typical horizontal collimator sequences used in detecting superlattice peaks were 15$^{\prime}$(30$^{\prime}$)-30$^{\prime}$-30$^{\prime}$-80$^{\prime}$ and 80$^{\prime}$-40$^{\prime}$-40$^{\prime}$-80$^{\prime}$ for TOPAN and TAS-1, respectively. 
Additionally, pyrolytic graphite filters were inserted both up and down stream of the sample position in order to eliminate the higher-order reflected beams. 
Each sample was mounted with the ({\it h} {\it k} 0) plane parallel to the scattering plane. 
Followed by a previous work ~\cite{Fujita_01_1}, we denote the reciprocal space by the high-temperature tetragonal (HTT) phase with {\it I}4/{\it mmm} symmetry. 

To characterize the crystal structure we first examined nuclear Bragg peaks for each sample. 
In all samples, the (0 1 0) reflection, which is not allowed in the LTO phase, was observed at lowest temperature corresponding to either the LTT or low-temperature less-orthorhombic (LTLO) phase with {\it Pccn} symmetry. 
The double (0 1 0) peak from twinned orthorhombic domains in {\it x} = 0.06,0.075 and 0.085 samples indicates the LTLO phase, while the single (0 1 0) peak in {\it x} = 0.05 sample is consistent with the LTT phase. 

As seen in Figs. 1(a) and (b), both IC superlattice peaks from CDW and SDW orders are clearly observed for the {\it x} = 0.05 sample in the LTT phase consistent with results for the LNSCO system ~\cite{Tranquada95}. 
However, in the case of the {\it x} = 0.085 sample in the LTLO phase peak intensity from the CDW order is dramatically reduced while well-defined peak from SDW order is observed (see Figs. 1(c) and (d)). 
Note that the finite intensity of CDW peak in Fig. 2(b) plotted at {\it x} = 0.085 was obtained by a measurement with higher counting rate compared to the case shown in Fig. 1(c).
The vertical scales in Fig. 1 are normalized by taking the sample volume estimated from phonon intensities into account. 
Hence relative peak intensities between the two samples can be compared quantitatively. 
It is also noted that the magnetic IC peak intensity in La$_{1.875}$Ba$_{0.075}$Sr$_{0.05}$CuO$_{4}$ \cite{Kimura99_2,KimuraPrive} is approximately six times stronger than that of La$_{1.88}$Sr$_{0.12}$CuO$_{4}$ ({\it T}$_{c}$ = 31.5 K) ~\cite{Kimura99}. 
Therefore both static CDW and SDW are well stabilized in the LTT phase where superconductivity is strongly suppressed. 

In Fig. 2, volume-corrected peak intensities integrated within the planes are shown for the (0 1 0) reflection and the CDW and SDW peaks as a function of Sr concentration. 
We observed the same shift of both static SDW and CDW peaks to a low symmetric positions in the LTLO system ~\cite{Fujita_unpublish} as previously observed for SDW peaks in LTO systems ~\cite{YoungLee99,Kimura00}. 
We note these peak-shifts do not affect the integrated values for the peak intensities. 
As shown in Figs. 2(a) and (b), the Sr-substitution similarly reduces both intensities of the (0 1 0) reflection and the CDW order. 
In fact, no well-defined CDW peak is observed in the LTO phase where (0 1 0) reflection is not allowed. 
On the other hand, the substantial intensity from the SDW order remains even in the LTO phase. 
These results strongly suggest the stability of CDW order more strongly depends on the crystal structure than the SDW order. 
Furthermore, a direct relation of the CDW order with the suppression of {\it T}$_{c}$ becomes clear 
\linebreak
\begin{figure}[htbp]
\vspace{-8mm}
\centerline{\epsfxsize=2.2in\epsfbox{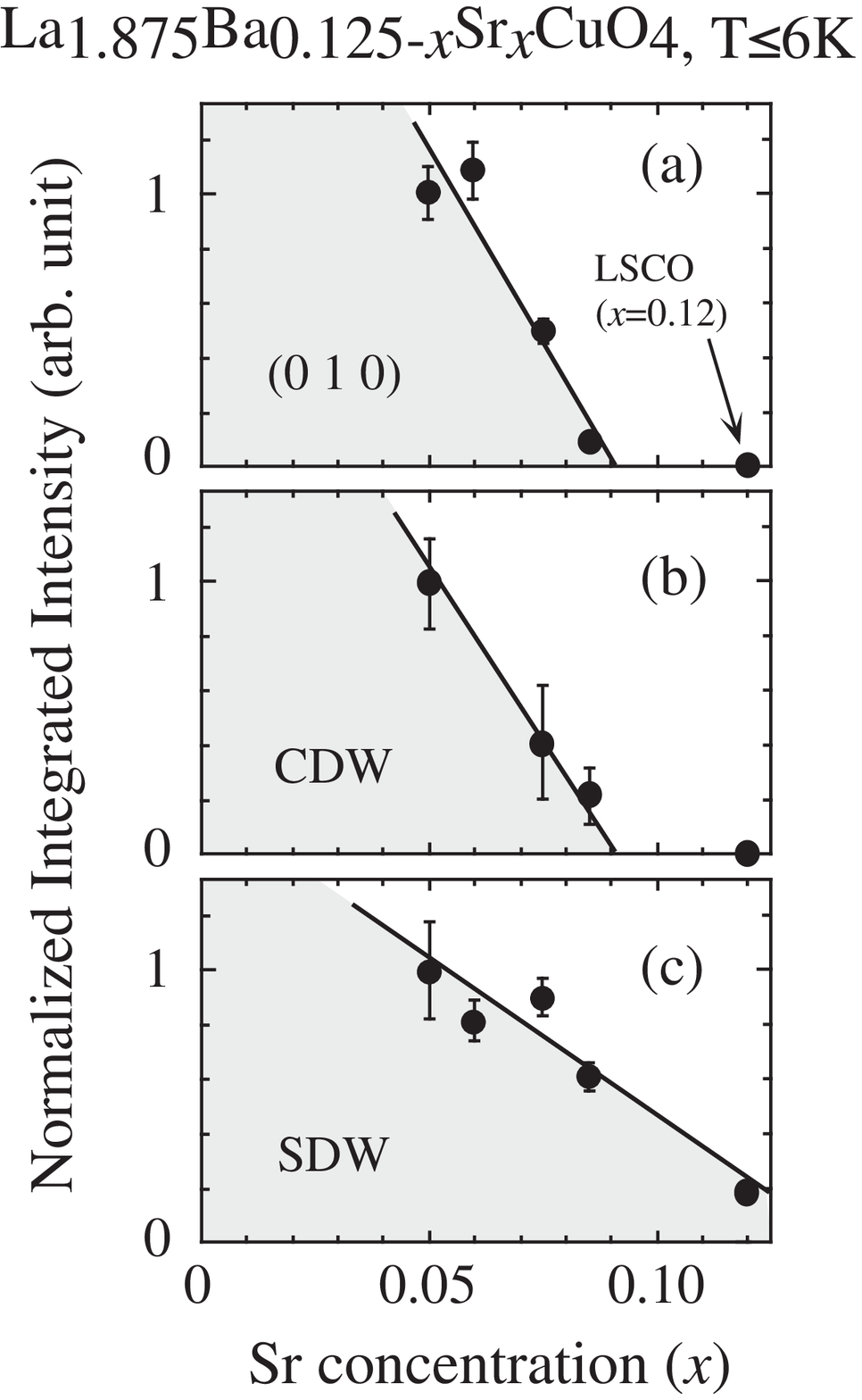}}
\vspace{0.5cm}
\caption
{
Peak intensity integrated within the plane for (a) (0 1 0), (b) CDW and (c) SDW peaks as a function of Sr concentration. Result for La$_{1.88}$Sr$_{0.12}$CuO$_{4}$ is plotted as a reference [18,19]. Solid lines are guides to the eye.
}
\end{figure}
\vspace{0cm}
\noindent
when the {\it x}-dependence of {\it T}$_{c}$ in Fig. 4 is mapped on Fig. 2(b) \cite{Fujita_01_2,Goka_unpublish}. 
In other words, the strong instability of superconductivity is considered to be triggered by the charge localization rather than by magnetic pair-braking. 

Next, we introduce the second new result, that is, coincident appearance of both CDW and SDW orders. 
In Fig. 3, the temperature dependence of the order parameter is shown for the (0 1 0), CDW and SDW peak intensities with the {\it x} = 0.05 sample. 
Intensities are normalized by the data at 1.4 K after subtracting the background taken at a higher temperature. 
Both the CDW and SDW peak intensities were found to begin appearing at the structural transition temperature {\it T}$_{d2}$ (= 37 K) with similar temperature dependence to that of the order parameter of the (0 1 0). 
The coincident appearance of the CDW and SDW orders below {\it T}$_{d2}$ is also observed for the {\it x} = 0.075 ({\it T}$_{d2}$ = 32 K) and 0.085 ({\it T}$_{d2}$ = 30 K) samples. 
This behavior, however, is qualitatively different from that found in the LNSCO system, in which the SDW ordering temperature ({\it T}$_{m}$) is lower than {\it T}$_{d2}$ or the CDW ordering temperature. 
One possible explanation for such distinct behavior between the two systems will be addressed later in this paper. 

In Fig. 4, we present the superconducting \cite{Fujita_01_2,Goka_unpublish} 
and structural phase diagram as a function of Sr concentration. 
{\it T}$_{c}$ of La$_{1.88}$Sr$_{0.12}$CuO$_{4}$ is also plotted as a reference ~\cite{Kimura99}. 
Shaded area corresponds to the LTT or LTLO phase. 
In the figure, we set the ground state 
\linebreak
\begin{figure}[htbp]
\vspace{-8mm}
\centerline{\epsfxsize=2.08in\epsfbox{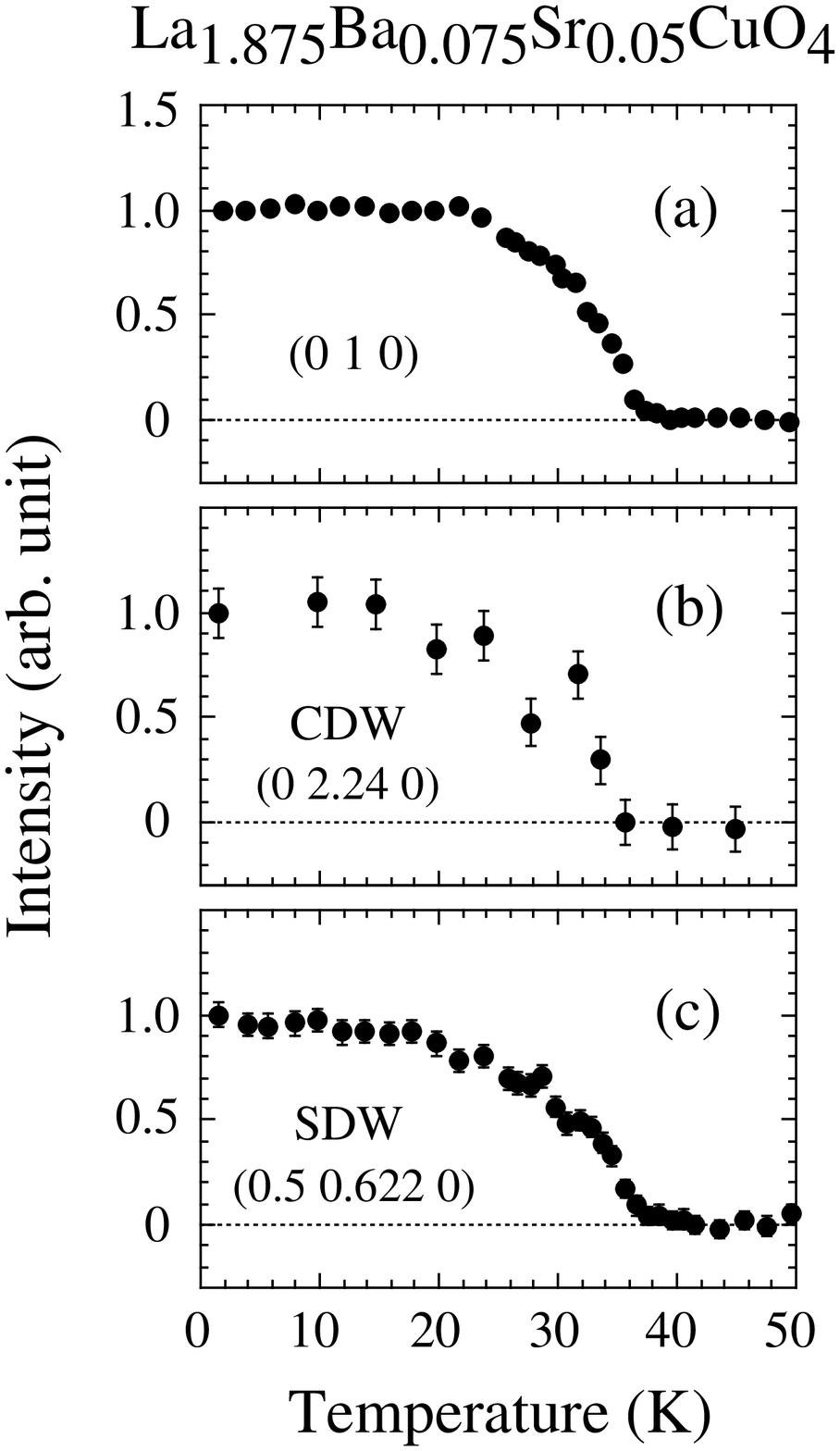}}
\vspace{0.5cm}
\caption
{
Temperature dependencies of (a) (0 1 0), (b) CDW and (c) SDW superlattice peak intensities. 
}
\end{figure}
\noindent
LTLO-LTO phase boundary at {\it x}$_{c}$ $\sim$0.09 with the extrapolation of (0 1 0) peak intensity shown in Fig. 2(a). 
An abrupt drop in {\it T}$_{c}$ with decreasing {\it x} is then found around the phase boundary. 
That is, {\it T}$_{c}$ in the LTO phase is independent of {\it x}, while in the LTT/LTLO phase {\it T}$_{c}$ varies for 0.05$\leq${\it x}$\leq$0.085 where {\it T}$_{d2}$ changes. 
This result also suggests the close interplay between superconductivity and the crystal structure. 

Previously other groups already reported close relation between the superconductivity and the crystal structure. Dabrowski {\it et al}. found the strong correlation between {\it T}$_{c}$ and the tilt angle of the CuO$_{6}$ octahedron or the amplitude of the corrugated lattice potential of CuO$_{2}$ planes ~\cite{Dabrowski}. 
Furthermore, in both thin films and pressure-applied 2-1-4 compounds with flat CuO$_{2}$ planes the disappearance of the 1/8-anomaly and the substantial increment of {\it T}$_{c}$ are reported \cite{Sato00_1,Sato00_2,Goko99,Nakamura00}. 
Such interplay is easily interpreted through the stripe model or from viewpoint of the stability of CDW order. 
In terms of the stripe model, as CDW order on the flat CuO$_{2}$ plane is free from the pinning potential, {\it T}$_{c}$ of the orthorhombic phase also should be restored. 

Finally, we discuss the possible reason for the different ordering sequence between LNSCO and LBSCO systems. 
When {\it T}$_{d2}$ is high enough, CDW order is expected to appear at {\it T}$_{ch}$$^{\prime}$ with decreasing temperature followed by SDW order at {\it T}$_{m}$$^{\prime}$ as the schematic ordering sequence is shown in Fig. 5(a). 
Then, if {\it T}$_{d2}$ is lowered, since static CDW and SDW are more stabilized in the LTT than in the LTO phase, these orders can be controlled by 
\linebreak
\begin{figure}[htbp]
\vspace{-8mm}
\centerline{\epsfxsize=2.4in\epsfbox{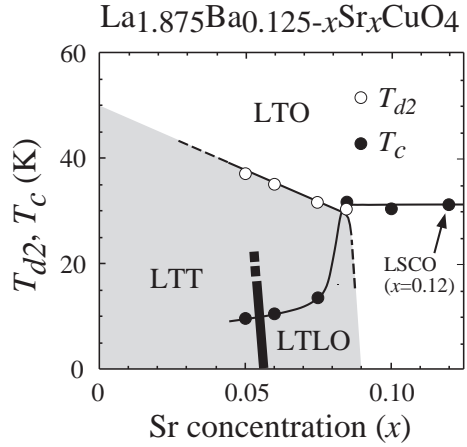}}
\vspace{0.5cm}
\caption
{
Phase diagram of {\it T}$_{d2}$ (open circles) and {\it T}$_{c}$ (closed circles.) for the La$_{1.875}$Ba$_{0.125-x}$Sr$_{x}$CuO$_{4}$ system. {\it T}$_{c}$$^{\prime}$s are taken from Refs. 12, 13 and 20. Thick lines correspond to LTT-LTLO phase boundary at low temperature. Solid and dashed lines are guides to the eye. 
}
\end{figure}
\noindent
the structural transition at {\it T}$_{d2}$. 
LNSCO and LBSCO systems correspond to the cases shown in Figs. 5(b) and (c), respectively: only the CDW ordering for the former and both CDW and SDW orderings for the latter are triggered by the phase transition to the LTT phase. 
Therefore the coincident appearance of both orders in LBSCO system is originated from the lower {\it T}$_{d2}$ than that in LNSCO system. 

In conclusion, we found in the 1/8-hole doped LBSCO system the strong structural effect on the CDW order. 
The order is stabilized in the LTT/LTLO phase and thereby severely suppresses the superconductivity, while no well-defined CDW order is observed in the LTO phase where the suppression of {\it T}$_{c}$ is small. 
Most of these results are explained through the stripe model. 
However, the absence of the CDW order in the LTO phase with remaining static/quasi-static SDW is difficult to 
\linebreak
\begin{figure}[htbp]
\vspace{-0mm}
\centerline{\epsfxsize=2.4in\epsfbox{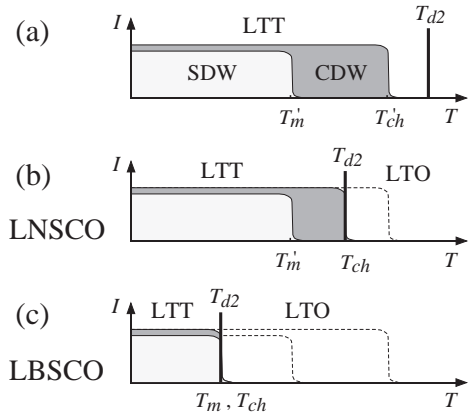}}
\vspace{0.5cm}
\caption
{
Schematic ordering sequences for the cases of (a) {\it T}$_{d2}$ $>$ {\it T}$_{ch}$$^{\prime}$, (b) {\it T}$_{m}$$^{\prime}$ $\leq$ {\it T}$_{d2}$ $\leq$ {\it T}$_{ch}$$^{\prime}$ and (c) {\it T}$_{d2}$ $<$ {\it T}$_{m}$$^{\prime}$. 
}
\end{figure}
\noindent
understand by the simple stripe model. 
More comprehensive study on the charge ordering in the LTO phase is highly required. 

We would like to thank H. Kimura, Y. S. Lee, G. Shirane and J. M. Tranquada for stimulating discussions. This work was supported in part by the Japanese Ministry of Education, Culture, Sports, Science and Technology, Grant-in-Aid for Scientific Research on Priority Areas (Novel Quantum Phenomena in Transition Metal Oxides), 12046239, 2000, for Scientific Research (A), 10304026, 2000, for Encouragement of Young Scientists, 13740216, 2001 and for Creative Scientific Research (13NP0201) "Collaboratory on Electron Correlations - Toward a New Research Network between Physics and Chemistry -", by the Japan Science and Technology Corporation, the Core Research for Evolutional Science and Technology Project (CREST).



\end{document}